\documentclass[aps,twocolumn, superscriptaddress,showpacs,pra,amssymb, amsmath]{revtex4}
\usepackage{amsmath,latexsym,amssymb}
\usepackage{graphicx}
\usepackage{amsmath}
\usepackage{latexsym}
\usepackage{amssymb}
\usepackage{bm}

\begin{document}

\title{Bell inequality with an arbitrary 
number of settings and its applications}

\author{Koji Nagata}
\affiliation{Instytut Fizyki Teoretycznej i Astrofizyki,
Uniwersytet Gda\'nski, PL-80-952 Gda\'nsk, Poland}
\affiliation{National Institute of Information and Communications Technology,
4-2-1 Nukuikita, Koganei, Tokyo 184-8795, Japan}

\author{Wies{\l}aw Laskowski}
\affiliation{Instytut Fizyki Teoretycznej i Astrofizyki,
Uniwersytet Gda\'nski, PL-80-952 Gda\'nsk, Poland}

\author{Tomasz Paterek}
\affiliation{Instytut Fizyki Teoretycznej i Astrofizyki,
Uniwersytet Gda\'nski, PL-80-952 Gda\'nsk, Poland}
\affiliation{The Erwin Schr\"odinger International Institute for Mathematical Physics,
Boltzmanngasse 9, A-1090 Vienna, Austria}

\date{\today}

\begin{abstract}
Based on a geometrical argument introduced by \.Zukowski,
a new multisetting Bell inequality is derived,
for the scenario in which many parties make measurements
on two-level systems.
This generalizes and unifies some previous results.
Moreover, a necessary and sufficient condition
for the violation of this inequality is presented.
It turns out that the class of non-separable states
which do not admit local realistic description
is extended when compared to the two-setting inequalities.
However, supporting the conjecture of Peres,
quantum states with positive partial transposes 
with respect to all subsystems
do not violate the inequality.
Additionally, we follow a general link between Bell inequalities and
communication complexity problems,
and present a quantum protocol linked with the inequality,
which outperforms the best classical protocol.
\end{abstract}

\maketitle

\section{Introduction}

Any theory based on classical concepts,
such as locality and realism, 
predicts bounds on the correlations
between measurement outcomes obtained in space-separation \cite{BELL}.
These bounds are known as Bell inequalities (see \cite{REVIEWS} for reviews).
Profoundly, the correlations measured
on certain quantum states violate Bell inequalities,
implying incompatibility between the quantum and classical worldviews.
Which are these non-classical states of quantum mechanics?
Here, we present a tool which allows one to extend the class of non-classical states,
and gives further evidence that there may exist many-particle entangled states
whose correlations admit a local realistic description.

Despite their fundamental role,
with the emergence of quantum information \cite{NC},
Bell inequalities have found practical applications.
Quantum advantages of certain protocols,
like quantum cryptography \cite{CRYPTOGRAPHY} or quantum communication complexity \cite{BRUKNER_PRL},
are linked with Bell inequalities.
Thus, new inequalities lead to new schemes.
As an example, we present communication complexity problem
associated with the new multisetting inequality.

Specifically,
based on a geometrical argument by \.Zukowski \cite{ZUKOWSKI_PLA},
a Bell inequality for many observers,
each choosing between arbitrary number of dichotomic 
observables, is derived.
Many previously known inequalities are special cases of this new inequality,
e.g. Clauser-Horne-Shimony-Holt inequality \cite{CHSH}
or tight two-setting inequalities \cite{MERMIN}.
The new inequalities
are maximally violated by the Greenberger-Horne-Zeilinger (GHZ) states \cite{GHZ}.
Many other states violate them,
including the states which satisfy two-settings inequalities \cite{ZBLW}
and bound entangled states \cite{DUR}.
This is shown using the necessary and sufficient condition
for the violation of the inequalities.
Finally, it is proven that the Bell operator
has only two non-vanishing eigenvalues
which correspond to the GHZ states,
and thus has a very simple form.
This form is utilized to show that quantum states
with positive partial transposes \cite{PPT} with respect 
to all subsystems 
(in general the necessary but not sufficient condition for entanglement \cite{HORODECKI})
do not violate the new inequalities.
This is further supporting evidence 
for a conjecture by Peres 
that positivity of partial transposes 
could lead us to the existence of a local realistic model \cite{PERES}.

The paper is organized as follows.
In section II we present the multisetting inequality.
In section III 
the necessary and sufficient condition for a violation of the inequality is derived,
and examples of non-classical states are given.
Next, we support the conjecture by Peres in section IV,
and follow the link with communication complexity problems
in section V.
Section VI summarizes this paper.

\section{Multisetting Bell Inequalities}

Consider $N$ separated parties making measurements on two-level systems.
Each party can choose one of $M$ dichotomic, of values $\pm 1$, observables.
In this scenario parties can measure $M^N$ correlations $E_{m_1...m_N}$,
where the index $m_n=0,...,M-1$ denotes the setting of the $n$th observer.
A general Bell expression, which involves these correlations
with some coefficients $c_{m_1...m_N}$,
can be written as:
\begin{equation}
\sum_{m_1, ..., m_N =0}^{M-1} c_{m_1...m_N} E_{m_1...m_N} = \vec C \cdot \vec E.
\label{GENERAL_BELL}
\end{equation}
In what follows we assume certain form of coefficients $c_{m_1...m_N}$,
and compute local realistic bound as a maximum of a scalar product
$|\vec C \cdot \vec E^{LR}|$. 
The components of vector
$\vec E^{LR}$ have the usual form:
\begin{equation}
E_{m_1...,m_N}^{LR} = \int d \lambda \rho(\lambda) I_{m_1}^1(\lambda)...I_{m_N}^N(\lambda),
\label{E_LR}
\end{equation}
where $\lambda$ denotes a set of hidden variables, $\rho(\lambda)$ their distribution,
and $I_{m_n}^{n}(\lambda) = \pm 1$ the predetermined result of $n$th observer under setting $m_n$.
The quantum prediction for the Bell expression (\ref{GENERAL_BELL})
is given by a scalar product of $\vec C \cdot \vec E^{QM}$.
The components of $\vec E^{QM}$, 
according to quantum theory,
are given by:
\begin{equation}
E_{m_1...m_N}^{QM} = {\rm Tr}\left( \rho \vec m_1 \cdot \vec \sigma^1 \otimes ... \otimes \vec m_N \cdot \vec \sigma^N \right),
\end{equation}
where $\rho$ is a density operator (general quantum state),
$\vec \sigma^n = (\sigma_x^n,\sigma_y^n,\sigma_z^n)$ is a vector of local Pauli operators
for $n$th observer,
and $\vec m_n$ denotes a normalized vector 
which parameterizes observable $m_n$ for the $n$th party.

Assume that local settings are parameterized by a single angle:
$\phi_{m_n}^n$.
In the quantum picture we restrict observable vectors $\vec m_n$
to lie in the equatorial plane:
\begin{equation}
\vec m_n \cdot \vec \sigma^n = \cos\phi_{m_n}^n \sigma_x^n + \sin\phi_{m_n}^n \sigma_y^n.
\end{equation}
Take the coefficients $c_{m_1...m_N}$
in a form:
\begin{equation}
c_{m_1...m_N} = \cos(\phi_{m_1}^1 + ... + \phi_{m_N}^N),
\end{equation}
with the angles given by:
\begin{equation}
\phi_{m_n}^n = \frac{\pi}{M} m_n + \frac{\pi}{2MN} \eta.
\label{ANGLES}
\end{equation}
The number $\eta=1,2$ is fixed for a given experimental
situation, i.e. $M$ and $N$, and equals:
\begin{equation}
\eta = [M+1]_2 [N]_2 + 1,
\label{ETA}
\end{equation}
where $[x]_2$ stands for $x$ modulo $2$.
The local realistic bound is given by 
a maximal value of the scalar product $|\vec C \cdot \vec E^{LR}|$.
The maximum is attained for deterministic local realistic models,
as they correspond to the extremal points of a correlation polytope.
Thus, the following inequality appears:
\begin{eqnarray}
&&|\vec C \cdot \vec E^{LR}| \le \label{INEQ_DETER} \\
&&\max_{I^1_0, ..., I^N_{M-1} = \pm 1} \left\{ \sum_{m_1,...,m_N=0}^{M-1} \! \! \! \! \! \!
\cos(\phi_{m_1}^1 + ... + \phi_{m_N}^N)
I_{m_1}^1...I_{m_N}^N \right\},
\nonumber
\end{eqnarray}
where we have shortened the notation $I_{m_n}^n \equiv I_{m_n}^n(\lambda)$.
Since 
$\cos(\phi_{m_1}^1 + ... + \phi_{m_N}^N) = {\rm Re} \left( \prod_{n=1}^N \exp{(i \phi_{m_n}^n)} \right)$
and the predetermined results, $I_{m_n}^n = \pm 1$, are real,
the right-hand side of this inequality can be written as:
\begin{equation}
\sum_{m_1,...,m_N=0}^{M-1} 
{\rm Re} \left( \prod_{n=1}^N \exp{(i \phi_{m_n}^n)} I_{m_n}^n \right).
\end{equation}
Moreover, since inequality (\ref{INEQ_DETER}) involves
the sum of all possible products of local results respectively
multiplied by the cosines of all possible sums of local angles,
the right-hand side can be further reduced to
involve the product of sums:
\begin{equation}
{\rm Re} \left( \prod_{n=1}^N \sum_{m_n=0}^{M-1}\exp{(i \phi_{m_n}^n)} I_{m_n}^n \right).
\end{equation}
Inserting the angles (\ref{ANGLES}) into this expression results in:
\begin{equation}
{\rm Re} \left( \exp{(i \frac{\pi}{2M} \eta)} 
\prod_{n=1}^N \sum_{m_n=0}^{M-1}\exp{(i \frac{\pi}{M}m_n)} I_{m_n}^n \right),
\label{RE_ANGLES}
\end{equation}
where the factor $\exp{(i \frac{\pi}{2M} \eta)}$
comes from the term $\frac{\pi}{2MN} \eta$ in (\ref{ANGLES}),
which is the same for all parties.  

One can decompose a complex number given by the sum in (\ref{RE_ANGLES})
into its modulus $R_n$, and phase $\Phi_n$:
\begin{equation}
\sum_{m_n=0}^{M-1}\exp{(i \frac{\pi}{M}m_n)} I_{m_n}^n = R_n e^{i\Phi_n}.
\label{VECTOR}
\end{equation}
We maximize the length of this vector on 
the complex plane.
The length of the sum of any two complex numbers
$|z_1 + z_2|^2$ is given by the law of cosines
as $|z_1|^2+|z_2|^2 + 2 |z_1| |z_2| \cos \varphi$,
where $\varphi$ is the angle between the corresponding vectors.
To maximize the length of the sum
one should choose the summands as close as possible
to each other.
Since in our case all vectors being summed
are rotated by multiples of $\frac{\pi}{M}$
from each other, the simplest optimal choice
is to put all $I_{m_n}^n = 1$.
In this case one has:
\begin{equation}
R_n^{\max} = \left| \sum_{m_n=0}^{M-1}\exp{(i \frac{\pi}{M}m_n)} \right|
= \left| \frac{2}{1-\exp{(i\frac{\pi}{M})}} \right|,
\end{equation}
where the last equality follows from the finite sum of numbers in the
geometric progression (any term in the sum is given by the preceding term multiplied by $e^{i
\pi/M}$).
The denominator inside the modulus can be transformed to
$\exp{(i\frac{\pi}{2M})} \left[ \exp{(-i\frac{\pi}{2M})} - \exp{(i\frac{\pi}{2M})} \right]$,
which reduces to $- 2i \exp{(i\frac{\pi}{2M})} \sin\left(\frac{\pi}{2M}\right)$.
Finally, the maximal length reads:
\begin{equation}
R_n^{\max} = \frac{1}{\sin\left(\frac{\pi}{2M}\right)},
\end{equation}
where the modulus is no longer needed 
since the argument of sine is small.
Moreover, since the local results for each party can be chosen independently, 
the maximal length $R_n^{\max}$ does not depend on particular $n$, i.e. $R_n^{\max} = R^{\max}$.

Since $R^{\max}$ is a positive real number 
its $N$th power can be put
to multiply the real part in (\ref{RE_ANGLES}),
and one finds $|\vec C \cdot \vec E^{LR}|$ to be bounded by:
\begin{equation}
|\vec C \cdot \vec E^{LR}| \le
\left[ \sin \left(\frac{\pi}{2M} \right) \right]^{-N} \! \! \! \! 
\cos \left( \frac{\pi}{2M} \eta + \Phi_1 + ... + \Phi_N \right),
\end{equation}
where the cosine comes from the phases of the sums in (\ref{RE_ANGLES}).
These phases can be found from the definition (\ref{VECTOR}).
As only vectors rotated by a multiple of $\frac{\pi}{M}$ are summed 
(or subtracted) in (\ref{VECTOR}), each phase $\Phi_n$ can acquire
only a restricted set of values.
Namely:
\begin{equation}
\Phi_n = 
\Bigg\{
\begin{array}{lc}
\frac{\pi}{2M} + \frac{\pi}{M}k & \textrm{ for } M \textrm{ even}, \\
 & \\
\frac{\pi}{M}k & \textrm{ for } M \textrm{ odd},
\end{array}
\end{equation}
with $k=0,...,2 M-1$,
i.e. for $M$ even, $\Phi_n$ is an odd multiple of $\frac{\pi}{2 M}$;
and for $M$ odd, $\Phi_n$ is an even multiple of $\frac{\pi}{2 M}$.
Thus, the sum $\Phi_1 + ... + \Phi_N$
is an even multiple of $\frac{\pi}{2 M}$,
except for $M$ even and $N$ odd.
Keeping in mind the definition of $\eta$, given in (\ref{ETA}),
one finds the argument of
$\cos \left( \frac{\pi}{2M}\eta + \Phi_1 + ... + \Phi_N \right)$
is always odd multiple of $\frac{\pi}{2 M}$,
which implies the maximum value of the cosine is equal to
$\cos \left(\frac{\pi}{2M} \right)$.
Finally, the multisetting Bell inequality reads:
\begin{equation}
|\vec C \cdot \vec E^{LR}| \le \left[ \sin \left(\frac{\pi}{2M}\right) \right]^{-N} \cos \left(\frac{\pi}{2M} \right).
\label{INEQUALITY}
\end{equation}
This inequality,
when reduced to two parties choosing between two settings each,
recovers the famous Clauser-Horne-Shimony-Holt inequality \cite{CHSH}.
For higher number of parties, still choosing between two observables,
it reduces to tight two-setting inequalities \cite{MERMIN}.
When $N$ observers choose between three observables 
the inequalities of \.Zukowski and Kaszlikowski are obtained \cite{ZUK_KASZ},
and for continuous range of settings ($M \to \infty$) it recovers the inequality of \.Zukowski \cite{ZUKOWSKI_PLA}.

\section{Quantum Violations}

In this section we present a Bell operator
associated with the inequality (\ref{INEQUALITY}).
Next, it is used to 
derive the necessary and sufficient condition
for the violation of the inequality.
Using this condition we recover already known results
and present some new ones.

The form of the coefficients $c_{m_1...m_N} = \cos(\phi_{m_1}^1 + ... + \phi_{m_N}^N)$
we have chosen is exactly the same
as the quantum correlation function
$E_{m_1...m_N}^{GHZ} = \cos(\phi_{m_1}^1 + ... + \phi_{m_N}^N)$
for the Greenberger-Horne-Zeilinger state:
\begin{equation}
|\psi^+ \rangle = \frac{1}{\sqrt{2}}
\Big[ |0\rangle_1 ... |0\rangle_N + |1\rangle_1 ... |1\rangle_N \Big],
\end{equation}
where the vectors $|0 \rangle_n$ and $|1 \rangle_n$
are the eigenstates of local $\sigma_z^n$ operator of the $n$th party.
For this state the two vectors $\vec C$ and $\vec E^{GHZ}$ are equal (thus parallel),
which means that 
the state $|\psi^+ \rangle$ maximally violates inequality (\ref{INEQUALITY}).
The value of the left hand side of (\ref{INEQUALITY}) is given
by the scalar product of $\vec E^{GHZ}$ with itself:
\begin{equation}
\vec E^{GHZ} \cdot \vec E^{GHZ}
= \sum_{m_1,...,m_N=0}^{M-1} \cos^2(\phi_{m_1}^1 + ... + \phi_{m_N}^N).
\label{COS_SQUARE}
\end{equation}
Using the trigonometric identity $\cos^2 \alpha = \frac{1}{2}(1+\cos2\alpha)$
one can rewrite this expression into the form:
\begin{equation}
\vec E^{GHZ} \cdot \vec E^{GHZ}
= \frac{1}{2}M^N + \frac{1}{2}\sum_{m_1,...,m_N=0}^{M-1} 
\! \! \! \! \! \! \! \! \cos[2(\phi_{m_1}^1 + ... + \phi_{m_N}^N)].
\end{equation}
As before, the second term can be written as a real part of a complex number.
Putting the values of angles (\ref{ANGLES})
one arrives at:
\begin{equation}
\frac{1}{2} {\rm Re}
\left( \exp{(i \frac{\pi}{M} \eta)} \prod_{n=1}^N \sum_{m_n=0}^{M-1}\exp{(i \frac{2\pi}{M}m_n)}\right).
\label{COMPLEX_ROOTS}
\end{equation}
Note that $e^{i\frac{2\pi}{M}}$ is a primitive complex $M$th root of unity.
Since all complex roots of unity
sum up to zero
the above expression vanishes,
and a maximal quantum value of the left hand side of (\ref{INEQUALITY}) 
equals:
\begin{equation}
\vec E^{GHZ} \cdot \vec E^{GHZ} = \frac{1}{2}M^N.
\end{equation}
If instead of $| \psi^+ \rangle$ one chooses
the state $| \psi^- \rangle = \frac{1}{\sqrt{2}}
[ |0\rangle_1 ... |0\rangle_N - |1\rangle_1 ... |1\rangle_N ]$,
for which the correlation function is given by
$E_{m_1...m_N}^{GHZ-} = - \cos(\phi_{m_1}^1 + ... + \phi_{m_N}^N)$,
one arrives at a minimal value of the Bell expression, equal to $-\frac{1}{2}M^N$,
as the vectors $\vec C$ and $\vec E^{GHZ-}$
are exactly opposite.
Since we take a modulus in the Bell expression,
both states lead to the same violation.

The Bell operator associated with the Bell expression (\ref{INEQUALITY})
is defined as:
\begin{equation}
\mathcal{B'} \equiv \! \! \! \! \! \! \! 
\sum_{m_1...m_N=0}^{M-1} \! \! \! \! \! \! c_{m_1...m_N} 
\vec m_1 \cdot \vec \sigma^1 \otimes ... \otimes \vec m_N \cdot \vec \sigma^N.
\label{B}
\end{equation}
Its average in the quantum state $\rho$
is equal to the quantum prediction of the Bell expression, for this state.
We shall prove that it has only two
eigenvalues $\pm \frac{1}{2}M^N$,
and thus is of the simple form:
\begin{equation}
\mathcal{B} \equiv
\mathcal{B}(N,M) = \frac{1}{2}M^N \left[ |\psi^+ \rangle \langle \psi^+ | - 
|\psi^- \rangle \langle \psi^- |\right].
\label{BELL_OPERATOR}
\end{equation}

Both operators $\mathcal{B}$ and $\mathcal{B'}$
are defined in the Hilbert-Schmidt space
with the trace scalar product.
To prove their equivalence one should
check if the conditions:
\begin{equation}
{\rm Tr}(\mathcal{B'} \mathcal{B}) = 
{\rm Tr}(\mathcal{B} \mathcal{B}) = 
{\rm Tr}(\mathcal{B'} \mathcal{B'}),
\label{TRACES}
\end{equation}
are satisfied.
Geometrically speaking, these conditions
mean that
the ``length'' and ``direction''
of the operators are the same.

The trace ${\rm Tr}(\mathcal{B'} \mathcal{B})$
involves the traces 
${\rm Tr}\left(|\psi^{\pm} \rangle \langle \psi^{\pm} | \vec m_1 \cdot \vec \sigma^1 \otimes ... \otimes \vec m_N \cdot \vec \sigma^N \right)$,
which are the quantum correlation functions
(averages of the product of local observables) for the GHZ states,
and thus are given by $\pm \cos(\phi_{m_1}^1 + ... + \phi_{m_N}^N)$.
Their difference doubles the cosine,
which is then multiplied by the same cosine
coming from the coefficients $c_{m_1...m_N}$.
Thus the main trace takes the form:
\begin{equation}
{\rm Tr}(\mathcal{B} \mathcal{B'}) =
M^N \! \! \! \! \! \! \sum_{m_1...m_N=0}^{M-1} \! \! \! \! \!
\cos^2(\phi_{m_1}^1 + ... + \phi_{m_N}^N) 
= \frac{1}{2} M^{2N},
\end{equation}
where the last equality sign
follows from the considerations below Eq. (\ref{COS_SQUARE}).

The middle trace of (\ref{TRACES})
is given by ${\rm Tr}(\mathcal{B} \mathcal{B}) = \frac{1}{2} M^{2N}$,
which directly follows from the
orthonormality of the states $| \psi^{\pm} \rangle$.

The last trace of (\ref{TRACES}) is more involved.
Inserting decomposition (\ref{B})
into ${\rm Tr}(\mathcal{B'} \mathcal{B'})$ gives:
\begin{eqnarray*}
&& \sum_{\substack{
m_1...m_N, \\
m_1'...m_N'=0}}^{M-1} 
\! \! \! \! \!
\cos(\phi_{m_1}^1 + ... + \phi_{m_N}^N) \cos(\phi_{m_1'}^1 + ... + \phi_{m_N'}^N) \\
&& \times {\rm Tr}[(\vec m_1 \cdot \vec \sigma^1)(\vec m_1' \cdot \vec \sigma^1) ]...
{\rm Tr}[(\vec m_N \cdot \vec \sigma^N)(\vec m_N' \cdot \vec \sigma^N) ]
\end{eqnarray*}
The local traces are given by:
\begin{equation}
{\rm Tr}[(\vec m_n \cdot \vec \sigma^n)(\vec m_n' \cdot \vec \sigma^n) ]
= 2 \vec m_n \cdot \vec m_n' = 2 \cos(\phi_{m_n}^n - \phi_{m_n'}^n).
\end{equation}
Thus, the factor $2^N$ appears in front of the sums.
We write all the cosines (of sums and differences) in terms of individual
angles, insert these decompositions into ${\rm Tr}(\mathcal{B'} \mathcal{B'})$,
and perform all the multiplications.
Note that whenever the final product term involves
at least one expression like
$\cos\phi_{m_n}^n \sin \phi_{m_n}^n = \frac{1}{2}
\sin(2 \phi_{m_n}^n)$
(or for the primed angles)
its contribution to the trace vanish after the summations
[for the reasons discussed in Eq. (\ref{COMPLEX_ROOTS})].
Moreover, in the decomposition of 
$\cos(\phi_{m_n}^n - \phi_{m_n'}^n) = \cos\phi_{m_n}^n \cos\phi_{m_n'}^n +
\sin\phi_{m_n}^n \sin\phi_{m_n'}^n$
only the products of the same trigonometric functions appear.
In order to contribute to the trace
they must be multiplied by again the same functions.
Since the decompositions of cosines of sums
only differ in angles (primed or unprimed)
and not in the individual trigonometric functions,
the only contributing terms come from the product
of exactly the same individual trigonometric functions
in the decomposition of $\cos(\phi_{m_1}^1 + ... + \phi_{m_N}^N)$ 
and $\cos(\phi_{m_1'}^1 + ... + \phi_{m_N'}^N)$.
There are $2^{N-1}$ such products,
as many as the number of terms in the decomposition
of $\cos(\phi_{m_1}^1 + ... + \phi_{m_N}^N)$.
Each product involves $2N$ squared 
individual trigonometric functions.
Each of these functions can be written in terms
of cosines of a double angle, e.g. $\sin^2 \phi_{m_n}^n =
\frac{1}{2}(1-\cos(2\phi_{m_n}^n))$,
and the last cosine does not contribute to the sum
[again due to (\ref{COMPLEX_ROOTS})].
Finally the trace reads:
\begin{equation}
{\rm Tr}(\mathcal{B'} \mathcal{B'}) = 2^N
\! \! \! \! \!
\sum_{ \substack{ 
m_1...m_N, \\ m_1'...m_N'=0}}^{M-1} 
\! \! \! \! \!
2^{N-1} \frac{1}{2^{2N}} = \frac{1}{2} M^{2N}.
\end{equation}
Thus, equations (\ref{TRACES})
are all satisfied, i.e.
both operators $\mathcal{B}$ and $\mathcal{B'}$ are equal.
Only the states which 
have contributions in the subspace spanned by $| \psi^{\pm} \rangle$
can violate the inequality (\ref{INEQUALITY}).

\emph{Necessary and sufficient condition for the violation
of the inequality}.
The expected quantum value of the Bell expression,
using Bell operator, reads:
\begin{equation}
{\rm Tr}(\mathcal{B}(N,M)\rho) = 
\frac{M^N}{2} \left[ {\rm Tr}(|\psi^+ \rangle \langle \psi^+ |\rho) - 
{\rm Tr}(|\psi^- \rangle \langle \psi^- | \rho)\right].
\label{AVERAGED_BELL_OP}
\end{equation}
The violation condition is obtained after maximization,
for a given state, over the position of the $xy$ plane, 
in which the observables lie.

An arbitrary state (density operator) of $N$ qubits
can be decomposed using local Pauli operators as:
\begin{equation}
\rho = \frac{1}{2^N} \sum_{\mu_1...\mu_N=0}^{3} T_{\mu_1...\mu_N} 
\sigma_{\mu_1} \otimes ... \otimes \sigma_{\mu_N},
\end{equation}
where the set of averages $T_{\mu_1...\mu_N} = {\rm Tr}[\rho(\sigma_{\mu_1} \otimes ... \otimes \sigma_{\mu_N})]$
forms the so-called correlation tensor.
The correlation tensors of the projectors 
$|\psi^\pm \rangle \langle \psi^\pm |$ are denoted by $T_{\nu_1...\nu_N}^{\pm}$.
Using the linearity of the trace operation
and the fact that the trace of the tensor product
is given by the product of local traces,
one can write ${\rm Tr}(|\psi^\pm \rangle \langle \psi^\pm | \rho)$
in terms of correlation tensors:
\begin{equation}
\frac{1}{2^{2N}} \! \! \! \!
\sum_{\substack{
\mu_1...\mu_N,\\ 
\nu_1...\nu_N=0}}^3 
\! \! \! \! 
T_{\nu_1...\nu_N}^{\pm} T_{\mu_1...\mu_N}
{\rm Tr}(\sigma_{\mu_1} \sigma_{\nu_1})...{\rm Tr}(\sigma_{\mu_N} \sigma_{\nu_N}). \nonumber
\end{equation}
Since each of the $N$ local traces ${\rm Tr}(\sigma_{\mu_n} \sigma_{\nu_n}) = 2 \delta_{\mu_n \nu_n}$,
the global trace is given by:
\begin{equation}
{\rm Tr}(|\psi^\pm \rangle \langle \psi^\pm | \rho)
= \frac{1}{2^{N}}  \sum_{\mu_1...\mu_N=0}^3
T_{\mu_1...\mu_N}^{\pm} T_{\mu_1...\mu_N}.
\label{TRACING}
\end{equation}
The nonvanishing correlation tensor components of the GHZ states $| \psi^{\pm} \rangle$
are the same in the $z$ plane: $T_{z...z}^{\pm} = 1$ for $N$ even;
and are exactly opposite in the $xy$ plane:
$T_{i_1...i_N}^{+} = - T_{i_1...i_N}^{-} = (-1)^\xi$
with $2\xi$ indices equal to $y$ and all remaining equal to $x$.
Inserting the traces (\ref{TRACING})
into the averaged Bell operator (\ref{AVERAGED_BELL_OP})
one finds that the components in the $z$ plane
cancel out, and components in the $xy$ plane
double themselves.
Finally, 
the necessary and sufficient condition to satisfy the inequality is given by:
\begin{equation}
\left( \frac{M}{2}\right)^N \max \sum_{i_1...i_N \in I_\xi} (-1)^{\xi} T_{i_1...i_N} \le 
 B_{LR}(N,M),
\label{NS}
\end{equation}
where the maximization is performed over the choice of local
coordinate systems, 
$I_\xi$ includes all sets of indices $i_1...i_N$ with 2$\xi$
indices equal to $y$ and the rest equal to $x$,
and 
\begin{equation}
B_{LR}(N,M) = \left[ \sin \left(\frac{\pi}{2M}\right) \right]^{-N} \cos \left(\frac{\pi}{2M} \right)
\end{equation}
denotes the local realistic bound.

We now present examples of states,
which violate the new inequality.
As a measure of violation, $V(N,M)$, 
we take the average (quantum) value of the Bell operator
in a given state,
divided by the local realistic bound:
\begin{equation}
V(N,M) = \frac{\langle \mathcal{B}(N,M) \rangle_{\rho}}{B_{LR}(N,M)}.
\end{equation}

\begin{figure}
\begin{center}
\includegraphics[width=0.5\textwidth]{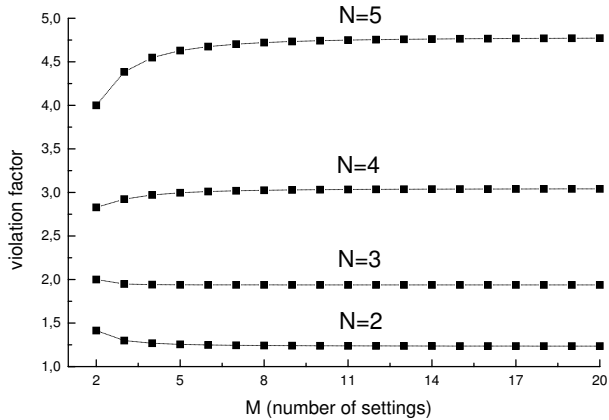}
\end{center}
\caption{Violation factor as a function of number of local settings, $M$,
for the $N$-qubit GHZ state.}
\label{M_PLOT}
\end{figure}

\emph{GHZ state}.
First, let us simply consider $| \psi^{\pm} \rangle$.
For the case of two settings per side
one recovers previously known results \cite{MERMIN,WW_BI,ZB}:
\begin{equation}
V(N,2) = 2^{(N-1)/2}.
\end{equation}
For three settings per side
the result of \.Zukowski and Kaszlikowski is obtained \cite{ZUK_KASZ}:
\begin{equation}
V(N,3) = \frac{1}{\sqrt{3}}\left(\frac{3}{2} \right)^{N}.
\end{equation}
For the continuous range of settings one recovers \cite{ZUKOWSKI_PLA}:
\begin{equation}
V(N,\infty) = \frac{1}{2}\left(\frac{\pi}{2} \right)^{N}.
\end{equation}
In the intermediate (unexplored before) regime one has:
\begin{equation}
V(N,M) = \frac{1}{2 \cos \left(\frac{\pi}{2M} \right)}\left(M \sin \left(\frac{\pi}{2M} \right) \right)^{N}.
\end{equation}
For a fixed number of parties $N > 3$ the violation
increases with the number of local settings.
Surprisingly, the inequality implies 
for the cases of $N=2$ and $N=3$
that the violation decreases when the number of local settings grows.
This behaviour is shown in the Fig. \ref{M_PLOT}.
The
violation
of local realism always grows
with increasing number of parties.

\emph{Generalized GHZ state.}
Consider the GHZ state with free real coefficients:
\begin{equation}
| \psi \rangle = 
\cos\alpha | 0 \rangle_1 ... | 0 \rangle_N + \sin\alpha | 1 \rangle_1 ... | 1 \rangle_N.
\end{equation}
Its correlation tensor in the $xy$ plane
has the following nonvanishing components:
$T_{x...x} = \sin2\alpha$,
and the components with 2$\xi$ indices equal to $y$
and the rest equal to $x$ take the value
of $(-1)^{\xi} \sin2\alpha$ 
(there are $2^{N-1}-1$ such components).
Thus, all $2^{N-1}$ terms contribute to the violation condition (\ref{NS}).
The violation factor is equal to $V(N,M)=\frac{M^N}{2 B_{LR}(N,M)} \sin2\alpha$.
For $N>3$ and $M>2$
the violation is bigger than the violation of standard two-setting inequalities \cite{ZB}.
Moreover, some of the states $| \psi \rangle$,
for small $\alpha$ and odd $N$, do not violate \emph{any} two-settings
correlation function Bell inequality \cite{ZBLW},
and violate the multisetting inequality.

\emph{Bound entangled state.}
Interestingly, the inequality can reveal non-classical
correlations of a bound entangled state introduced
by D\"ur \cite{DUR}:
\begin{equation}
\rho_N = \frac{1}{N+1} \left( |\phi \rangle \langle \phi| + \frac{1}{2} \sum_{k=1}^N (P_k + \tilde P_k) \right),
\end{equation}
with $|\phi \rangle = \frac{1}{\sqrt{2}} \left[ |0 \rangle_1 ... | 0 \rangle_N + e^{i \alpha_N} |1 \rangle_1 ... | 1\rangle_N \right]$
($\alpha_N$ is an arbitrary phase),
and $P_k$ being 
a projector on the state 
$|0 \rangle_1 ... |1 \rangle_k ... |0 \rangle_N$ with ``1'' 
on the $k$th position 
($\tilde P_k$ is obtained from $P_k$ after replacing ``0'' by ``1'' and vice versa).
As originally shown in \cite{DUR}
this state violates Mermin-Klyshko inequalities for $N \ge 8$.
The new inequality
predicts the violation factor of:
\begin{equation}
V(N,M) = \frac{1}{N+1} \frac{M^N ~\cos \alpha_N}{2B_{LR}(N,M)},
\label{viol-bes}
\end{equation}
which comes from the contribution
of the GHZ-like state $| \phi \rangle$
to the bound entangled state.
One can follow Ref. \cite{KASZLI} 
and change the Bell-operator (\ref{BELL_OPERATOR}) such that the state
$|\phi \rangle$ 
becomes its eigenstate. 
The new operator, $\tilde{\mathcal{B}} (N,M)$, 
is obtained after applying 
local unitary transformations
$U = |0\rangle \langle 0| + e^{i\alpha_N/N} |1\rangle \langle 1|$
to the operator (\ref{BELL_OPERATOR}),
i.e. 
$\tilde{\mathcal{B}} (N,M) = U^{\otimes N} \mathcal{B} U^{\dagger \otimes N}$.
The violation factor of the new inequality 
is higher than
(\ref{viol-bes}), and equal to:
\begin{equation}
\tilde{V}(N,M) = \frac{1}{N+1} \frac{M^N}{2B_{LR}(N,M)}.
\label{TILDE_VIS}
\end{equation}
If one sets $M=3$ it appears
that the number of parties sufficient
to see the violation (\ref{TILDE_VIS}) 
reduces to $N \ge 7$ \cite{KASZLI}.
On the other hand the result of \cite{ADITI}
shows that the infinite range of settings
further reduces the number of parties to $N \ge 6$.
Using the new inequality, 
$M=5$ settings per side suffice to
already violate local realism with $N \ge 6$ parties.

\section{Positive Partial Transpose}

In this section it is shown that all the states
with positive partial transpose with respect to all
subsystems satisfy the multisetting inequality (\ref{INEQUALITY}).
This result further supports the conjecture by Peres,
that all such states can admit local realistic description \cite{PERES}.
First we briefly review partial transpositions,
next present an inequality that all such states
must satisfy, and finally compare it with the Bell inequality (\ref{INEQUALITY}).

The partial transpose of an operator on a 
Hilbert space ${H}_1\otimes{H}_2$ is defined by:
\begin{eqnarray}
\left(\sum_l A^1_l\otimes A^2_l\right)^{T_1}=\sum_l{A_l^1}^{T} \otimes A_l^2,
\end{eqnarray}
where the superscript $T$ denotes transposition in the given basis.
The positivity of partial transpose is 
found to be a necessary condition for separability \cite{PPT,HORODECKI}. 
The operator obtained by the partial transpose of any separable state
is positive (PPT - positive partial transpose).
In the bipartite case of two qubits
or qubit-qutrit system,
the PPT criterion is also sufficient for separability.

In the multipartite case the situation complicates
as one can have many different partitions into set of particles,
for example four particle system $1234$ can be split 
e.g. into $12-34$ or $1-2-34$.
Suppose one splits $N$ particles into $p$ groups,
take as an example the split into three groups $1-2-34$.
The state is called $p$-PPT
if it has positive \emph{all possible} partial transposes.
Fortunately,
positivity of partial transpose 
with respect to certain set of subsystems
is the same as positivity 
with the respect to all remaining subsystems.
In the example one should check the positivity
of operator obtained after transposition of
subsystem $1$, next $2$, and finally $34$.

All the $p$-PPT states were recently shown 
to satisfy the following inequalities \cite{NAGATA}:
\begin{equation}
{\rm Tr}\left[ \left( |\psi^{\pm} \rangle \langle \psi^{\pm}| 
- (1-2^{2-p}) |\psi^{\mp} \rangle \langle \psi^{\mp}| \right) \rho \right] \le 2^{1-p},
\end{equation}
i.e. if $|\psi^+ \rangle$ appears in the first term within the trace,
$|\psi^- \rangle$ appears in the second term, and vice versa.
Omitting the positive factor 
$2^{2-p} {\rm Tr}\left( |\psi^{\mp} \rangle \langle \psi^{\mp}| \rho \right)$
one arrives at the Bell operator form:
\begin{equation}
\Big| {\rm Tr}\left[ \left( |\psi^+ \rangle \langle \psi^+| 
- |\psi^- \rangle \langle \psi^-| \right) \rho \right] \Big| \le 2^{1-p}.
\label{PPT}
\end{equation}
Following the conjecture by Peres let us put $p=N$.
We shall show that if inequality (\ref{PPT}) is satisfied,
with $p=N$, then also the Bell inequality (\ref{INEQUALITY}) is not violated.
Using the form of the Bell operator (\ref{BELL_OPERATOR}) the upper bound of the Bell inequality,
for $N$-PPT states, is found to read:
\begin{equation}
|{\rm Tr}(\mathcal{B}(N,M) \rho_{N-{\rm PPT}})| \le \left( \frac{M}{2} \right)^N,
\label{NPPT_BOUND}
\end{equation}
and it can never reach the local realistic bound $B_{LR}(M,N)$.
This is shown using the violation factor:
\begin{equation}
V_{N-{\rm PPT}}(N,M) = \left(\frac{M}{2} \right)^N \frac{[\sin \left(\frac{\pi}{2M} \right)]^N}{\cos \left(\frac{\pi}{2M} \right)}.
\end{equation}
Since $\sin \left(\frac{\pi}{2M} \right) \le \frac{\pi}{2M}$
and $\cos \left(\frac{\pi}{2M}\right) \ge \frac{1}{\sqrt{2}}$,
where we have put $M=2$ as a minimal amount of settings for which 
Bell inequality makes sense, the violation factor is bounded by
$V_{N-{\rm PPT}}(N,M) \le \sqrt{2} (\pi/4)^N$.
The simplest system on which one can perform partial transposes
consists of $N=2$ particles, 
thus $V_{N-{\rm PPT}}(N,M) \le \sqrt{2} (\pi/4)^2 \simeq 0.87$.
None of the $N$-PPT states violates the Bell inequality (\ref{INEQUALITY}).
It is worth mentioning that for $M=2$ setting case
the violation $V_{N-{\rm PPT}}(N,2) = 2^{(1-N)/2}$
confirms the results of Werner and Wolf \cite{WW_BI},
who gave the conjecture of Peres a sharp mathematical form \cite{WW_SHARP}.

\section{Communication Complexity}

Bell inequalities describe a performance of
quantum communication complexity protocols \cite{BRUKNER_PRL}.
In this section we follow this general link
and present communication complexity problems
associated with the inequality (\ref{INEQUALITY}).
It is proven that the quantum protocol
outperforms the best classical protocol
for arbitrary number of parties and observables.

In the communication complexity problems (CCP) 
one studies the information exchange between
participants \emph{locally} performing computations,
in order to accomplish a \emph{globally} defined task \cite{YAO}.
Let us focus on a variant of a CCP,
in which each of $N$ separated partners
receives arguments, $y_n=\pm 1$ and $x_n = 0,...,M-1$,
of some globally defined function, $\mathcal{F} \equiv \mathcal{F}(y_1,x_1,...,y_N,x_N)$.
The $y_n$ inputs are assumed to be randomly distributed,
and $x_n$ inputs can in general be distributed
according to a weight $\mathcal{W}(x_1,...,x_2)$.
The goal is to maximize the probability
that Alice arrives at the correct value of the function,
under the restriction that $N-1$ bits of overall
communication are allowed.
Before participants receive their inputs
they are allowed to do anything
from which they can derive benefit.
In particular, they can share some correlated strings of numbers
in the classical scenario or entangled states in the quantum case.

\emph{The problem.}
Following \cite{BRUKNER_PRL} one chooses for a task-function:
\begin{eqnarray}
\mathcal{F} &=& y_1...y_N {\rm Sign}[\cos(\phi_{x_1}^1 + ... + \phi_{x_N}^N)] = \pm 1,
\label{FF}
\end{eqnarray}
with the angles defined by (\ref{ANGLES}).
According to the angles definition 
the cosine can never be zero,
so the problem is well-defined for all $N$ and $M$.
Additionally, the $x_n$ inputs are distributed with the weight:
\begin{equation}
\mathcal{W}(x_1,...,x_2) = (1/\mathcal{N}) |\cos(\phi_{x_1}^1 + ... + \phi_{x_N}^N)|,
\label{WW}
\end{equation}
where the normalization factor is given by 
$\mathcal{N} = \sum_{x_1...x_N=0}^{M-1}|\cos(\phi_{x_1}^1 + ... + \phi_{x_N}^N)|$.
After the communication takes place,
if Alice misses some of the random variables $y_n$,
her ``answer'' can only be random.
Thus, in an optimal protocol 
each party must communicate one bit.
There are only two communication structures
which lead to a non-random answer:
(i) a star -- each party transmits one bit directly to Alice,
and (ii) a chain -- sequence of a peer-to-peer exchanges
with Alice at the end.
The task is to maximize the probability
of correct answer 
$\mathcal{A} \equiv \mathcal{A}(y_1,x_1,...,y_N,x_N)$.
Since both $\mathcal{A}$ and $\mathcal{F}$
are dichotomic variables this amounts in maximizing:
\begin{equation}
P_{{\rm correct}}
= \frac{1}{2^N}
\sum_{{\bf y},{\bf x}} \mathcal{W}(x_1,...,x_2)
P_{{\bf y},{\bf x}}(\mathcal{A} \mathcal{F} = 1),
\end{equation}
where $\frac{1}{2^N}$ describes (random) distribution of $y_n$'s,
and $P_{{\bf y},{\bf x}}(\mathcal{A} \mathcal{F} = 1)$
is a probability that $\mathcal{A} = \mathcal{F}$
for given inputs ${\bf y} \equiv (y_1,...,y_N)$ and ${\bf x} \equiv (x_1,...,x_N)$.
It is useful to express the last probability
in terms of an average value of a product $\langle \mathcal{A} \mathcal{F} \rangle_{{\bf y},{\bf x}}$,
i.e. 
$P_{{\bf y},{\bf x}}(\mathcal{A} \mathcal{F} = 1) = \frac{1}{2}[ 1 + \langle \mathcal{A} \mathcal{F} \rangle_{{\bf y},{\bf x}}]$.
Since $\mathcal{F}$ is independent of $\mathcal{A}$,
and for given inputs it is constant, one has
$P_{{\bf y},{\bf x}}(\mathcal{A} \mathcal{F} = 1) = \frac{1}{2}[ 1 + \mathcal{F} \langle \mathcal{A} \rangle_{{\bf y},{\bf x}}]$.
Finally the probability of correct answer reads 
$P_{{\rm correct}} = \frac{1}{2}[1 + (\mathcal{F},\mathcal{A})]$,
and it is in one-to-one correspondence with
a ``weighted'' scalar product (average success):
\begin{equation}
(\mathcal{F},\mathcal{A})
= \frac{1}{2^N}\sum_{{\bf y},{\bf x}} 
 \mathcal{W}(x_1,...,x_2) \mathcal{F} \langle \mathcal{A} \rangle_{{\bf y},{\bf x}}.
\end{equation}
Using the definitions (\ref{WW}) for 
$\mathcal{W}$ and (\ref{FF}) for $\mathcal{F}$
one gets:
\begin{equation}
(\mathcal{F},\mathcal{A})
= \frac{1}{2^N} \frac{1}{\mathcal{N}} \sum_{{\bf y},{\bf x}} 
 y_1...y_N \cos(\phi_{x_1}^1 + ... + \phi_{x_N}^N) \langle \mathcal{A} \rangle_{{\bf y},{\bf x}},
 \label{SUCCESS}
\end{equation}
with angles given by (\ref{ANGLES}).
We focus our attention on maximization of this quantity.

\emph{Classical scenario.}
In the \emph{best} classical protocol
each party locally computes a bit function $e_n = y_n f(x_n,\lambda)$,
with $f(x_n,\lambda) = \pm 1$,
where $\lambda$ denotes some previously shared
classical resources.
Next, the bit is
sent to Alice, who puts as an answer
the product
$\mathcal{A}_c = y_1 f(x_1,\lambda) e_2...e_N = y_1 ... y_N f(x_1,\lambda)...f(x_N,\lambda)$.
The same answer can be reached
in the chain strategy,
simply the $n$th party sends $e_n = y_n f(x_n,\lambda) e_{n-1}$.
For the given inputs the procedure is always the same,
i.e. $\langle \mathcal{A}_c \rangle_{{\bf y},{\bf x}} = \mathcal{A}_c$.
To prove the optimality of this protocol,
one follows the proof of Ref. \cite{EXP_CCP},
with the only difference that $x_n$ is a $M$-valued variable now.
This, however, does not invalidate any of the steps
of \cite{EXP_CCP}, and we will not repeat that proof.

Inserting the product form of $\mathcal{A}_c$
into the average success (\ref{SUCCESS}),
using the fact that $y_n^2=1$, and summing over all $y_n$'s
one obtains:
\begin{equation}
(\mathcal{F},\mathcal{A}_c) = 
\frac{1}{\mathcal{N}} \sum_{x_1...x_N=0}^{M-1} \! \! \! \! \! 
\cos(\phi_{x_1}^1 + ... + \phi_{x_N}^N)
f(x_1,\lambda)...f(x_N,\lambda),
\end{equation}
which has the same structure as local realistic expression (\ref{INEQ_DETER}).
Thus, the highest classically achievable average success
is given by a local realistic bound:
$\max (\mathcal{F},\mathcal{A}) = (1/\mathcal{N}) B_{LR}(N,M)$.

\emph{Quantum scenario.}
In the quantum case participants share a $N$-party entangled state $\rho$.
After receiving inputs each party measures $x_n$th observable on the state,
where the observables are enumerated as in the Bell inequality (\ref{INEQUALITY}).
This results in a measurement outcome, $f_n$.
Each party sends $e_n = y_n f_n$ to Alice,
who then puts as an answer a product
$\mathcal{A}_{q} = y_1...y_N f_1 ... f_N$.
For the given inputs the average answer
reads $\langle \mathcal{A}_{q} \rangle_{{\bf y},{\bf x}} = y_1...y_N  \langle f_1 ... f_N \rangle
= y_1...y_N E_{x_1...x_N}^{\rho}$,
and the maximal average success is given by a quantum bound of:
\begin{equation}
(\mathcal{F},\mathcal{A}_q) = 
\frac{1}{\mathcal{N}} \sum_{x_1...x_N=0}^{M-1} \! \! \! \! \! 
\cos(\phi_{x_1}^1 + ... + \phi_{x_N}^N)
E_{x_1...x_N}^{\rho}.
\end{equation}
The average advantage of quantum versus classical protocol
can be quantified by a factor $(\mathcal{F},\mathcal{A}_q)/(\mathcal{F},\mathcal{A}_c)$
which is equal to a violation factor, $V(N,M)$, introduced before.
Thus, 
all the states which violate the Bell inequality (including bound entangled state)
are a useful resource for the communication complexity task.
Optimally one should use the GHZ states $| \psi^{\pm} \rangle$,
as they maximally violate the inequality.

Alternatively, one can compare the probabilities of success, $P_{\textrm{correct}}$,
in quantum and 
classical case.
Clearly, one outperforms classical protocols
for every $N$ and every $M$. 
As an example, in Table \ref{TABLE_ADV}
we gather the ratios between quantum
and classical success probabilities
for small number of participants.

\begin{table}
\begin{tabular}{c | c c c c c}
$N \backslash M$  &   2     &   3     &   4   &   5   &   $\infty$   \\ 
\hline
2 & 1.1381       & 1.1196          &1.1009     &1.1002         &1.0909 \\
3 & 1.3333       & 1.2919          &1.2815     &1.2773         &1.2709 \\
4 & 1.3657       & 1.4395          &1.4038     &1.4258         &1.4192 \\
5 & 1.6000       & 1.5582          &1.5467     &1.5418         &1.5336  
\end{tabular}
\caption{The ration between probabilities of success in quantum and 
classical case $P_{\textrm{correct}}^{QM} / P_{\textrm{correct}}^{cl}$ for
the communication complexity problem with $N$ observers and $M$ settings.
Quantum protocol uses GHZ state.}
\label{TABLE_ADV}
\end{table}

One can ask about a CCP
with no random inputs $y_n$.
Since the numbers $x_n$ already represent $\lg M$ bits
of information, and only one bit can be communicated,
this looks like a plausible candidate for a quantum advantage.
However, in such a case
a classical answer cannot be put as a product
of outcomes of local computations (compare \cite{EXP_CCP}),
and thus there is no Bell inequality which would describe the best classical protocol.
Since classical performance of \emph{all} CCPs
which can lead to quantum advantage
is given by some Bell inequality \cite{BRUKNER_PRL},
the task without $y_n$'s cannot lead to quantum advantage.

\section{Summary}

We presented a multisetting Bell inequality,
which unifies and generalizes many previous results.
Examples of quantum states which violate the inequality were given.
It was also proven that all the states
with positive partial transposes with respect to all subsystems
cannot violate the inequality.
Finally, the states which violate it
were shown to reduce the communication complexity
of computation of certain globally defined function.
The Bell inequality presented
is the only inequality
which incorporates arbitrary number of settings
for arbitrary number of observers
making measurements on two-level systems, to date.

\acknowledgments

We thank M. {\.Z}ukowski for valuable discussions. 
W.L. and T.P. are supported by Foundation for Polish Science  
and MNiI Grant no. 1 P03B 049 27.
The work is part of the
VI-th EU Framework programme QAP (Qubit Applications) Contract No. 015848.


\begin{thebibliography}{9}

\bibitem{BELL}
J. S. Bell, 
Physics {\bf 1}, 195 (1964).

\bibitem{REVIEWS}
J. F. Clauser and A. Shimony,
Rep. Prog. Phys. {\bf 41}, 1881 (1978);
D. M. Greenberger, M. A. Horne, A. Shimony, and A. Zeilinger, Am. J. Phys. {\bf 58}, 1131 (1990); 
T. Paterek, W. Laskowski, and M. \.Zukowski,
Mod. Phys. Lett. A {\bf 21}, 111 (2006).

\bibitem{NC}
M. A. Nielsen and I. L. Chuang, 
{\it Quantum Computation and Quantum Information} 
(Cambridge University Press, Cambridge, England, 2000).

\bibitem{CRYPTOGRAPHY}
A. K. Ekert, Phys. Rev. Lett. {\bf 67}, 661 (1991);
V. Scarani and N. Gisin,
Phys. Rev. Lett. {\bf 87}, 117901 (2001);
A. Acin, N. Gisin, and V. Scarani,
Quant. Inf. Comp. {\bf 3}, 563 (2003).

\bibitem{BRUKNER_PRL}
{\v C}. Brukner, M. \.Zukowski, J.-W. Pan, and A. Zeilinger,
Phys. Rev. Lett. {\bf 92}, 127901 (2004).

\bibitem{ZUKOWSKI_PLA}
M. \.Zukowski,
Phys. Lett. A {\bf 177}, 290 (1993).

\bibitem{CHSH}
J. F. Clauser, M. A. Horne, A. Shimony, and R. A. Holt,
Phys. Rev. Lett. {\bf 23}, 880 (1969).

\bibitem{MERMIN}
N. D. Mermin,
Phys. Rev. Lett. {\bf 65}, 1838 (1990);
M. Ardehali,
Phys. Rev. A {\bf 46}, 5375 (1992);
A. V. Belinskii and D. N. Klyshko,
Phys. Usp. {\bf 36}, 653 (1993).


\bibitem{GHZ}
D. M. Greenberger, M. A. Horne, and A. Zeilinger,
in {\it Bell's Theorem, Quantum Theory and Conceptions of the Universe},
edited by M. Kafatos (Kluwer Academic, Dordrecht, The Netherlands, 
1989).

\bibitem{ZBLW}
M. \.Zukowski, {\v C}. Brukner, W. Laskowski, and M. Wie{\'s}niak,
Phys. Rev. Lett. {\bf 88}, 210402 (2002).

\bibitem{DUR}
W. D\"ur,
Phys. Rev. Lett. {\bf 87}, 230402 (2001).

\bibitem{PPT}
A. Peres, 
Phys. Rev. Lett. {\bf 77}, 1413 (1996);
M. Horodecki, P. Horodecki, and R. Horodecki, 
Phys. Lett. A {\bf 223}, 1 (1996).

\bibitem{HORODECKI}
M. Horodecki, P. Horodecki, and R. Horodecki, 
Phys. Rev. Lett. {\bf 80}, 5239 (1998).

\bibitem{PERES}
A. Peres,
Found. Phys, {\bf 29}, 589 (1999).

\bibitem{ZUK_KASZ}
M. \.Zukowski and D. Kaszlikowski, 
Phys. Rev. A {\bf 56}, R1682 (1997).

\bibitem{WW_BI}
R. F. Werner and M. M. Wolf, 
Phys. Rev. A {\bf 64}, 032112 (2001).

\bibitem{ZB}
M. \.Zukowski and \v{C}. Brukner, 
Phys. Rev. Lett. {\bf 88}, 210401 (2002).

\bibitem{KASZLI}
D. Kaszlikowski, L. C. Kwek, J. Chen, and C. H. Oh,
Phys. Rev. A {\bf 66}, 052309 (2002).

\bibitem{ADITI}
A. Sen (De), U. Sen, and M. \.Zukowski,
Phys. Rev. A {\bf 66}, 062318 (2002).

\bibitem{NAGATA}
K. Nagata,
Phys. Rev. A {\bf 66}, 064101 (2002).

\bibitem{WW_SHARP}
R. F. Werner and M. M. Wolf, 
Phys. Rev. A {\bf 61}, 062102 (2000).

\bibitem{YAO}
A. C.-C. Yao,
in {\it Proceedings of the 11th Annual ACM Symposium
on Theory of Computing}
(ACM Press, New York, 1979).

\bibitem{EXP_CCP}
P. Trojek, C. Schmid, M. Bourennane, {\v C}. Brukner, M. \.Zukowski, and H. Weinfurter,
Phys. Rev. A {\bf 72}, 050305(R) (2005).


\end{thebibliography}
\end{document}